\documentclass[epsf]{aa}

\usepackage{graphicx}
\begin{document}
\def\cd{cd$^{-1}$}
\def\cds{cd$^{-1}$\,}
\def\kms{km s$^{-1}$}
\def\kmss{km s$^{-1}$\,}
\def\td{$\theta^2$ Tau\,}

\title{Asteroseismology from space: the $\delta$ Scuti star
$\theta^2$ Tauri monitored by the WIRE satellite}

\author{E. Poretti\inst{1}, D.~Buzasi\inst{2}, R.~Laher\inst{3}, 
J.~Catanzarite\inst{4},T.~Conrow\inst{5}
    }
\offprints{E. Poretti}
\institute{Osservatorio Astronomico di Brera, Via Bianchi 46,
I-23807 Merate, Italy\\
\email{poretti@merate.mi.astro.it}
\and
Department of Physics, 2354 Fairchild Drive, US Air Force Academy,
CO 80840, USA
\and
SIRTF Science Center, California Institute of Technology, MS 314-6,
Pasadena, CA 91125, USA
\and 
Interferometry Science Center, California Institute of
Technology. MS 100-22, Pasadena, CA 91125, USA
\and
Infrared Processing and Analysis Center, California Insitute of
Technology, MS 100-22, Pasadena, CA 91125, USA
}

\date{Received date; accepted date }

\abstract { The bright variable star \td was monitored with the star camera
on the {\it Wide--Field Infrared Explorer} satellite. Twelve independent
frequencies were detected down to the 0.5 mmag amplitude level. 
Their reality was investigated by searching for them using
two different algorithms and by some internal checks: both procedures
strengthened our confidence in the results.
All the frequencies are in the range
10.8--14.6 \cd. The histogram of the frequency spacings
shows that 81\% are below 1.8~\cd; rotation may thus play a role
in the mode excitation. 
The fundamental radial mode is not observed, although it is expected to occur
in a region
where the noise level is very low (55$\mu$mag). The rms residual is about
two times lower than
that usually obtained from successful ground--based 
multisite campaigns. The comparison of the results of previous campaigns
with the new ones establishes the amplitude variability of some modes. 
}
\authorrunning{E. Poretti et al.}
\titlerunning{Asteroseismology of $\theta^2$ Tauri from space}
\maketitle

\begin{figure}
\resizebox{\hsize}{!}{\includegraphics{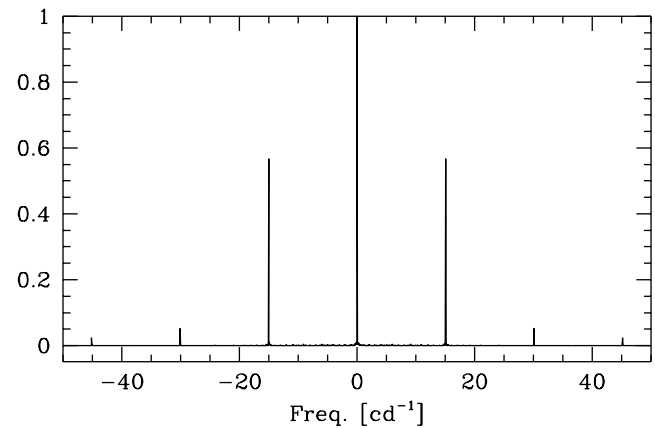}}
\caption[]{Window function of the \td  time series}
\label{sw}
\end{figure}

\section{Introduction}
\td, the brightest star of the Hyades, forms an optical pair
with $\theta^1$ Tau. The two stars are separated
by 5.6\arcmin\ and have a magnitude difference $\Delta V$=1.10 mag. \td is also
a spectroscopic 
binary, though direct detection of the secondary spectrum has been reported
only
by Peterson et al. (1993).
A reliable and recent solution of the orbit
has been proposed by Torres et al. (1997), though they were not able to
obtain
the radial velocity curve of the secondary. The primary  has a mass of
2.42$\pm$0.30 M\sun; the orbital period is  140.7282$\pm$0.0009~d 
and the orbit is highly eccentric, with $e$=0.727$\pm$0.005. 
Several photometric campaigns have clearly demonstrated
that the primary component of the \td system is a $\delta$ Scuti 
pulsating variable (Breger et al. 1989);
five terms having variable amplitude are reported by Li et al. (1997).

Soon after launch in March 1999, the primary science instrument onboard
the {\it Wide--Field Infrared Explorer} (WIRE) satellite failed due to loss
of coolant. However, it proved possible to begin an asteroseismology program
using the 52--mm aperture star camera.  A few bright stars were
monitored
with the 512x512 SITe CCD in a bandpass approximately equivalent to
$V+R$; further details about the orbit, the detector and the raw data
reduction can be found in Buzasi et al. (2000) and Buzasi (2000). 
The prospect of future space-based asteroseismology missions ({\sc COROT,
MONS, MOST}) 
has increased interest in
bright variable stars, a bit neglected in the past in favour of the 6--8
magnitude stars better-suited to differential photoelectric photometry
from the ground.
\td thus constituted both a good scientific target and a useful test for
asteroseismology from space.

\begin{figure*}
\resizebox{\hsize}{!}{\includegraphics{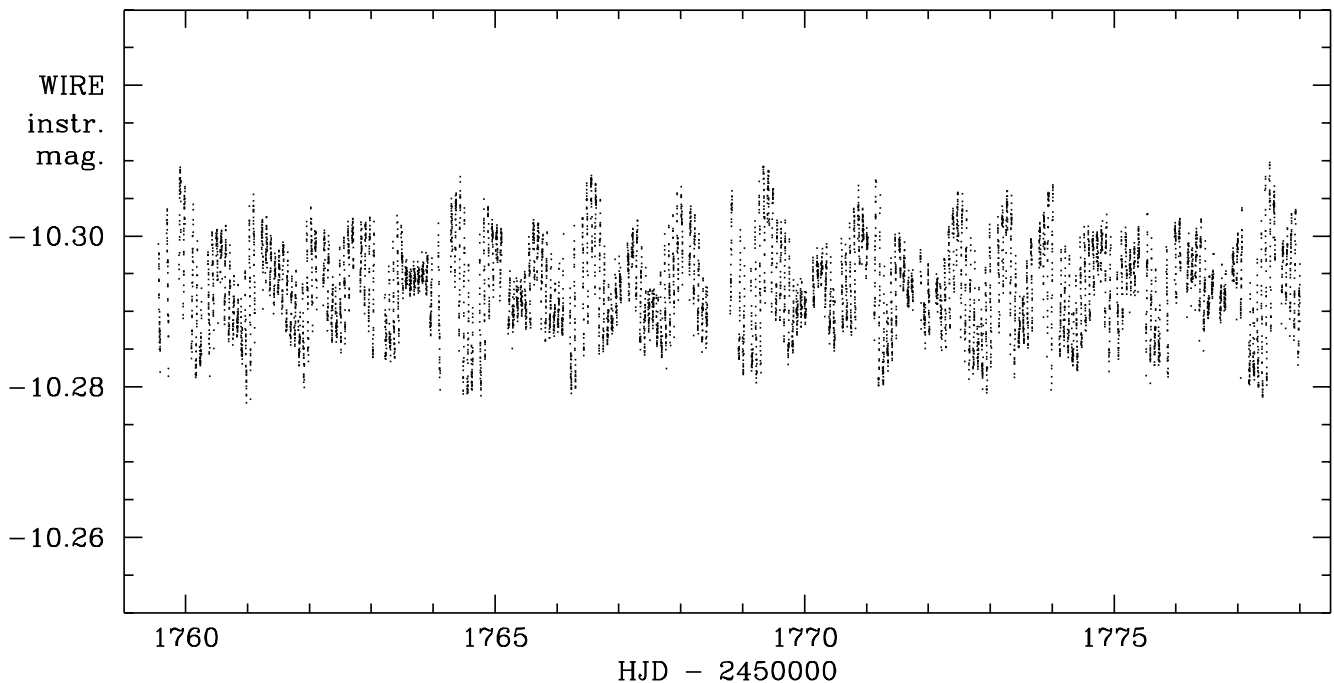}}
\caption[]{The light curve of \td  obtained with the 52--mm aperture
star camera on the WIRE satellite}
\label{curva}
\end{figure*}

\begin{figure*}
\resizebox{\hsize}{!}{\includegraphics{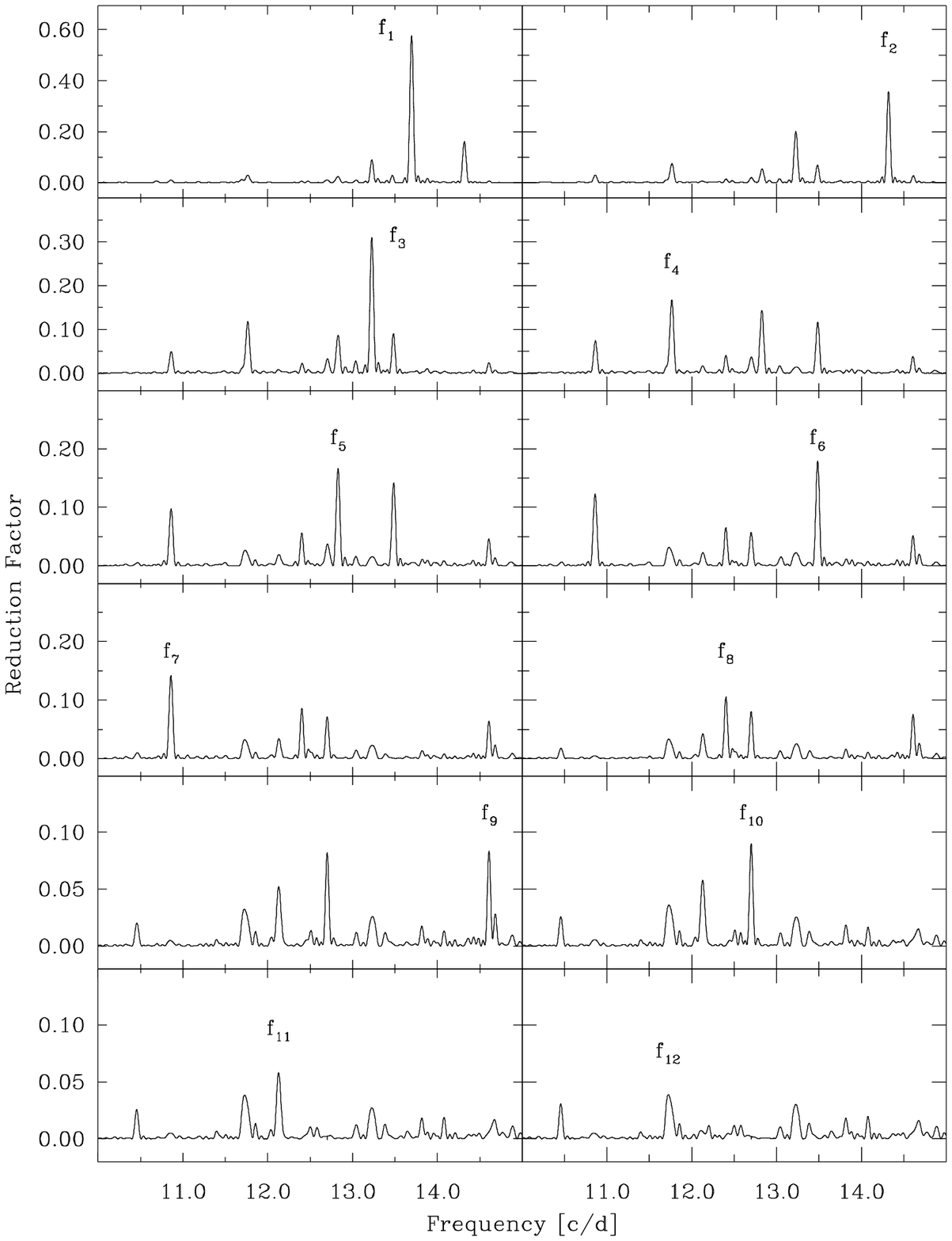}}
\caption[]{The least--squares power spectra of the WIRE observations of
\td. Each term is detected by considering the previously identified frequencies 
as known constituents in the least--squares solution.}
\label{lsq}
\end{figure*}

\section{Observations and data reduction}
\td was monitored from August 2 to 21, 2000;
the original dataset consists of 1,049,155 points. The typical
time interval between two consecutive measurements is 0.5 sec, 
resulting in  
an over-sampling of the light variability. 
As the luminosity
of \td varies by a negligible amount in one minute or so,
we grouped  the data in 60--sec bins, obtaining a dataset 
composed of 8958 normal points. 
The average value of the 8958 standard deviations 
yields us the
observational error on a single 0.5--sec integration, i.e. 5.9 mmag.
As the mean level is 13091.09 {\sc adu} and the gain is
15 $\mathrm~e^-/${\sc adu},
the resulting photon noise is 2.4 mmag on a single 0.5--sec integration.
As the observational error is more than twice the photon noise, it is
evident that other error sources are introduced by the
frame reading process. 
The  binning procedure we adopted reduced the error to about 0.5
mmag (standard error of the mean). 

The orbital period of the WIRE satellite is 5741~sec. 
The interruption of about 3480 sec (duty cycle 40\%) in each orbit 
simulates a night/day effect which originates the spectral window shown
in Fig.~\ref{sw}, dominated by the aliases at $\pm$15.05~\cd. 
Since the pulsational content of 
\td is expected to be very dense and confined to a small frequency 
range, it is a great advantage to have the alias region very far from 
that range. The data span an interval of about 18.5~d,
giving a frequency resolution of about 0.05~d$^{-1}$. 

The changing observational conditions (varying temperature, scattered light,
etc.)
caused by the satellite orbit, the jitter of the stellar image on the
detector
(a problem accentuated by the lack of a flat field) and the short duty cycle
are expected to introduce systematic deviations. As a
consequence, in our analysis we considered the frequencies we detected
at the orbital value (15.05 \cd), the duty cycle (26.19 \cds) and $f<1$~\cd
as spurious terms originated by these effects.   
The term at low frequency is also detected in the power spectrum of
the coordinates of the stellar centroid and hence no doubt is left 
as to its instrumental origin. 

Considering the long period of the spectroscopic binary and its small 
error bar, it is possible  to calculate the orbital phases of 
the WIRE run. We verified that it is located in the phase interval
0.50--0.64, where the light time correction is very small and
practically constant (see Figure~6 in Breger et al. 1989).
Therefore, we have not introduced such a correction.

Figure~\ref{curva} shows the light curve of \td derived from the 8958
averaged 60--sec bins: the three spurious periodicities 
at 15.05 \cd, 26.19 \cds and $f<1$~\cd have been removed. To do that,
after having obtained a first solution we  applied a least--squares
fit and then we removed the contribution of the three periodicities from the
data.
Instrumental magnitudes are $-2.5\log$({\sc adu}), where {\sc adu} are measured
in 0.5--sec intervals. Light variability and beating phenomena are
evident.

\section{Frequency content detection}
To detect the periodicities in the light curve, 
we used the least--squares iterative sine--wave
fitting approach (Vani\^cek 1971). It consists of the simultaneous
least--squares
fit of {\it n+1} sinusoids, where {\it n} represents the number
of the previously identified terms (known constituents, hereinafter
k.c.) and {\it n+1} is the the number of terms of the new trial 
solution. 
The reduction factor (i.e. how much the variance is reduced
by the $n+1$ frequency with respect to that calculated with the $n$
frequency solution) is given for each trial frequency in the range 
0--50 \cd.  This technique
is particularly suited to the case of multiperiodic light curves because
it does not require any prewhitening of the data. Indeed, the
amplitudes and phases of the terms previously identified are
recalculated when searching for the new one, i.e. only the frequency
values of the k.c.'s are kept constant. To avoid any possible
misidentification, we refined the frequency values by a non--linear least--squares
fit after the inclusion of a new term.  
Figure~\ref{lsq} shows the step--by--step detection by the
iterative sine--wave fittig procedure; the frequency values are listed in 
Table~\ref{erro}.

One of the most critical aspects in the signal detection concerns
the decision as to which peaks in the power spectrum can
be considered as intrinsic to the star. Due to the presence of nonrandom
errors
and because of observing gaps, the prediction of statistical false--alarm
tests give answers which are generally optimistic. 
To consider as real the peaks having $S/N~>~$4.0 
is a conservative  trade--off 
used by observers (Breger et al. 1993) and justified from a theoretical
point
of view (Kuschnig et al. 1997).  Therefore, we calculated 
the noise by averaging the amplitudes over a 10 \cds region centered
around the frequency under consideration; as sampling step, we used
1/20$\Delta$T, i.e. about 0.0025 \cd. The $S/N$ values calculated by
this way are listed in 
Table~\ref{erro}.

\begin{figure}
\resizebox{\hsize}{!}{\includegraphics{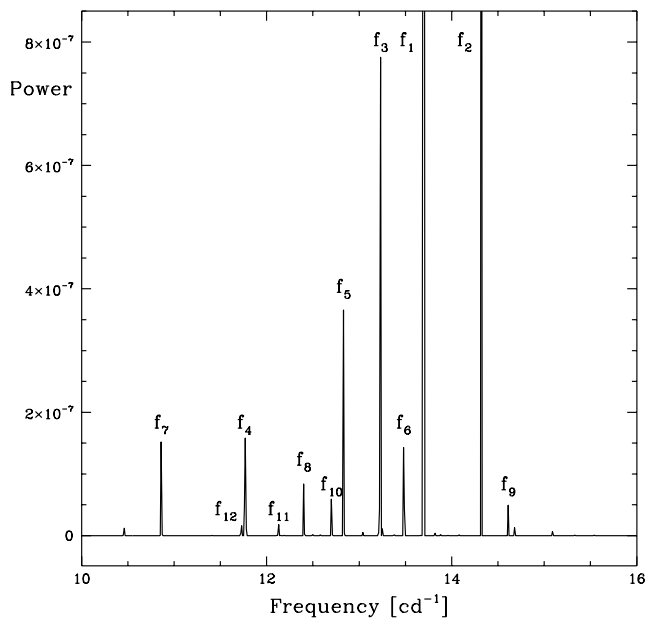}}
\caption[]{The {\sc clean}ed power spectrum of the WIRE observations of
\td. Note that the amplitudes of the $f_1$ and $f_2$ terms are off--scale.}
\label{clean}
\end{figure}

We duplicated the analysis by using the {\sc clean} algorithm (Roberts et
al. 1987): again we detected
the same frequencies (Figure~\ref{clean}). This is not surprising, 
since the spectral window
does not interact with the signal.   In turn, it means that in the
case of \td the sampling ensured by the WIRE monitoring has been very
effective.

To avoid supporting the frequency detection solely on a statistical
basis, we performed further checks.  
Looking closely at the frequency values shows
that, not considering the smallest amplitude term $f_{12}$=11.72 \cd, the shortest
separation is 13.69--13.48=0.21 \cd. That means it is possible to
perform the frequency analysis after first subdividing the dataset into two
subsets. These frequency analyses detected the same terms as in  the 
whole dataset. We calculated the least--squares fits on the two subsets,
taking care not to consider the unresolved, small amplitude $f_{12}$ term.
Moreover, Table~\ref{erro} 
reports the parameters of the least--squares fit on all the
data, on the first half of the data (4367 points before HJD 2451768.5) and
on the second half of the data (4591 points after  HJD 2451768.5). The average
error bars reported in Table~\ref{erro} are at the $2\sigma$ level.
Since we detected the same terms, in most of cases with the same amplitudes
and the same phases (within error bars), we are confident of the reality
of the frequencies listed in Table~\ref{erro}. Note also that the
$f_1$ and the $f_{10}$ terms are separated by 1.006 \cd; to resolve them
by single--site ground observation would prove a very hard task.
Formal errors (as derived from the least--squares fit) are of the
order of 10$^{-4}$~\cds for the highest amplitude terms and a few 10$^{-3}$~\cds
for the others.

\begin{figure*}
\resizebox{\hsize}{!}{\includegraphics{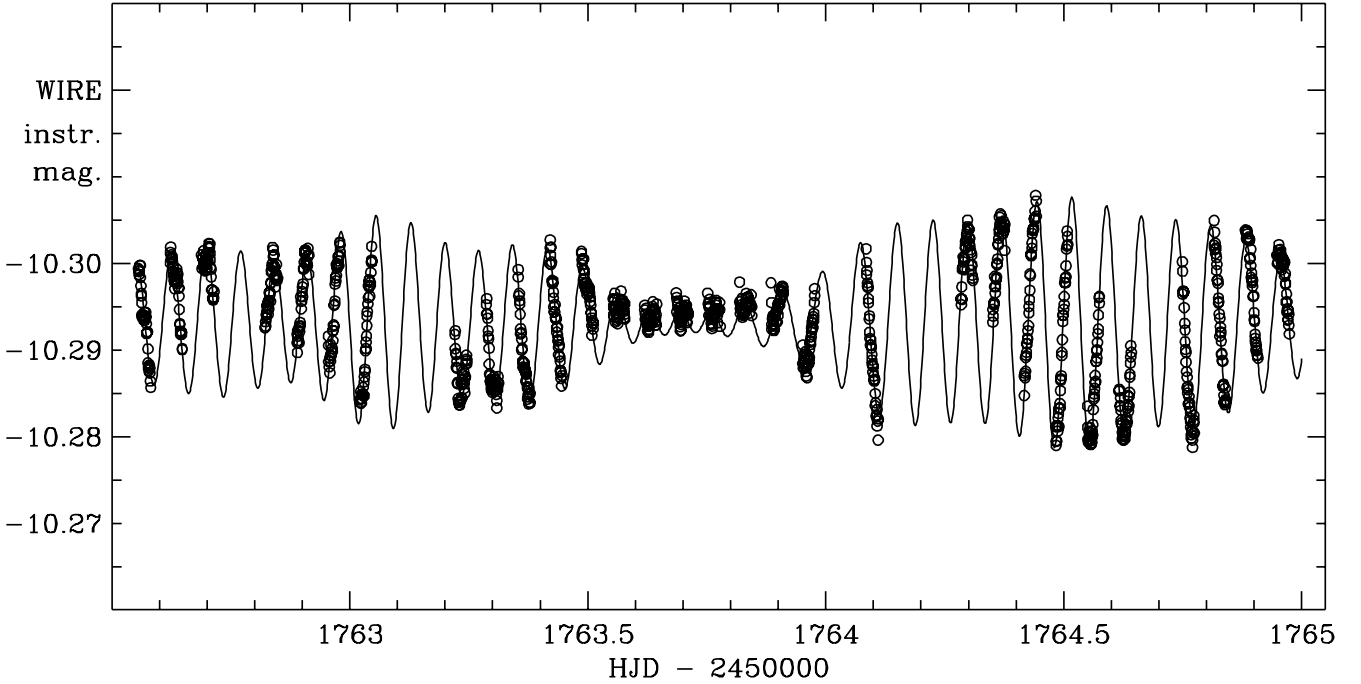}}
\caption[]{The fit of a part of the WIRE light curve
where beating is evident. The residual rms of the fit is
1.5 mmag.}
\label{fit}
\end{figure*}

We conclude that we have identified  12 independent terms in the WIRE light
curve of \td, down to 0.5 mmag half--amplitude level. This is the same 
limit reached on FG Vir,
the $\delta$ Sct star best studied from the ground. However, it should
be noted that for FG Vir this threshold was obtained by combining 3
different campaigns
(one of which was a multisite one involving six observatories for
40 d) spanning 10 years (Breger et al. 1998). 
It should also be noted that the residual rms of only 1.5 mmag is much smaller
than that obtained from multisite ground-based campaigns; it is the
limit sporadically reached in ground observatories located in very good 
photometric sites. 
To demonstrate the goodness of the least--squares solution,
Figure~\ref{fit} shows the 12--term fit of the normal points in a part of the 
WIRE light
curve where beating is evident: as it can be seen, the agreement
between observations and fit is excellent.
Figure~\ref{fit}  also shows the 
characteristic sampling of the WIRE time--series. 

\begin{center}
\begin{table*}
\caption{Terms detected in the light curve of \td and coefficients of the least--squares
fits. Phases are calculated respect with $T_0$=HJD 2451759.500} 
\begin{tabular}{l ccccc lcc lcc}
\hline
\multicolumn{1}{c}{}&\multicolumn{1}{c}{}&\multicolumn{1}{c}{}&
\multicolumn{1}{c}{}&
\multicolumn{2}{c}{All the data}& &
\multicolumn{2}{c}{First half of the data}& &
\multicolumn{2}{c}{Second half of the data} \\
\cline{5-6}\cline{8-9}\cline{11-12}
\multicolumn{1}{c}{}&\multicolumn{1}{c}{Frequency}&\multicolumn{1}{c}{Frequency}&
\multicolumn{1}{c}{S/N}&
\multicolumn{1}{c}{Amplitude}&\multicolumn{1}{c}{Phase}& &
\multicolumn{1}{c}{Amplitude}&\multicolumn{1}{c}{Phase}& &
\multicolumn{1}{c}{Amplitude}&\multicolumn{1}{c}{Phase} \\
%\cline{2-4}
 &\multicolumn{1}{c}{[\cd]}&\multicolumn{1}{c}{[$\mu$Hz]}& &
\multicolumn{1}{c}{[mmag]}&\multicolumn{1}{c}{[rad]} & &
\multicolumn{1}{c}{[mmag]}&\multicolumn{1}{c}{[rad]} & &
\multicolumn{1}{c}{[mmag]}&\multicolumn{1}{c}{[rad]} \\
\hline
\noalign{\smallskip}
$f_1$ &   13.6972 & 158.532 & 9.8  &  6.50 & 5.01 & & 6.56 & 5.01 & & 6.45 & 5.00    \\
$f_2$ &   14.3213 & 165.756 & 7.8  &  3.22 & 3.72 & & 3.20 & 3.73 & & 3.21 & 3.73 \\
$f_3$ &   13.2294 & 153.118 & 7.5  &  2.51 & 1.23 & & 2.87 & 1.21 & & 2.18 & 1.27 \\
$f_4$ &   11.7718 & 136.248 & 5.8  &  1.47 & 3.73 & & 1.22 & 3.98 & & 1.89 & 3.68   \\
$f_5$ &   12.8331 & 148.531 & 5.7  &  1.45 & 0.15 & & 1.46 & 0.17 & & 1.48 & 0.12\\
$f_6$ &   13.4870 & 156.099 & 6.1  &  1.32 & 5.12 & & 1.32 & 5.15 & & 1.30 & 5.02 \\
$f_7$ &   10.8613 & 125.709 & 5.8  &  1.07 & 5.68 & & 1.11 & 5.70 & & 0.99 & 5.61 \\
$f_8$ &   12.4043 & 143.568 & 5.3  &  0.83 & 4.98 & & 0.81 & 5.10 & & 0.92 & 4.95 \\
$f_9$ &   14.6104 & 169.102 & 4.7  &  0.72 & 2.94 & & 0.53 & 3.11 & & 0.91 & 2.86 \\
$f_{10}$& 12.7031 & 147.027 & 5.1  &  0.68 & 3.92 & & 0.54 & 3.96 & & 0.76 & 3.90 \\
$f_{11}$& 12.1274 & 140.363 & 4.5  &  0.55 & 1.40 & & 0.35 & 1.05 & & 0.73 & 1.51 \\
$f_{12}$& 11.7278 & 135.739 & 4.4  &  0.55 & 0.99 & & --   &  --  & &   -- & -- \\
\hline
\multicolumn{4}{c}{Average errors (2$\sigma$ level)}&$\pm$0.06&$\pm$0.09& &$\pm$0.09&$\pm$0.14& &
$\pm$0.16&$\pm$0.25\\
\multicolumn{4}{c}{Residual rms}&\multicolumn{2}{c}{1.45 mmag}& &\multicolumn{2}{c}{1.36
mmag}& &\multicolumn{2}{c}{1.49 mmag}\\
\hline
\end{tabular}
\label{erro}
\end{table*}
\end{center}

\section{Discussion}
\subsection{The presence of radial modes}
\begin{figure}
\resizebox{\hsize}{!}{\includegraphics{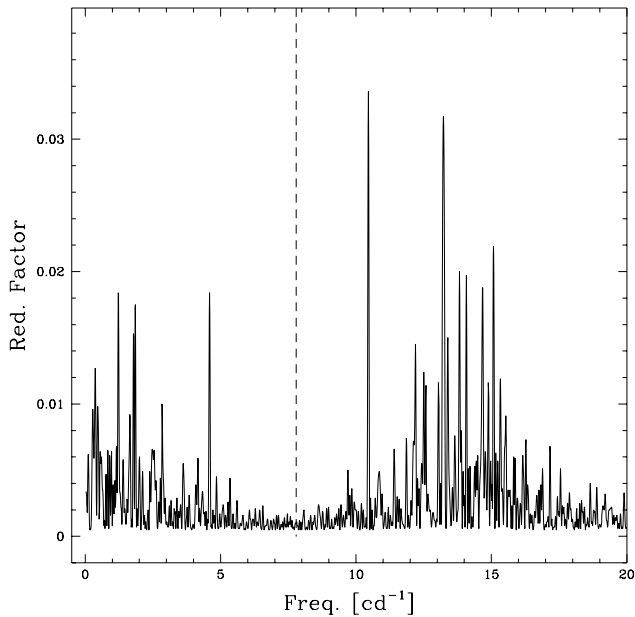}}
\caption[]{The residual least--squares power spectrum obtained considering 
the 12 terms as
k.c.'s; the predicted position of the unobserved radial
fundamental mode is indicated as a dashed line. The reduction factor
indicates how much the variance is reduced by the 13--th frequency
with respect to that calculated with the 12--frequency solution.}
\label{resi}
\end{figure}

Breger et al. (1989) used the following 
stellar parameters of \td:
$M_v=0.5, \log g=3.8, T_{\rm eff}=8200~K$. These values are
very similar to those adopted by Torres et al. (1997).
After applying the bolometric correction (Straizys \& Kuriliene 1981),
we can introduce them in the equation (Breger 2000)
\begin{equation}
\log Q = -6.456 -  \log \nu + 0.5 \log g + 0.1 M_{\rm bol} + \log T_{\rm eff}
\end{equation}
yielding $Q\nu$=0.257.
Therefore, we expect $\nu_0$=7.79, $\nu_1$=10.28,
$\nu_2$=12.85, $\nu_3$=15.12 and $\nu_4$=18.56 \cds for the fundamental
($F$, $Q$=0.033
d), first overtone ($1O$, 0.025 d), second overtone ($2O$, 0.020 d), third 
overtone ($3O$, 0.017 d) and
fourth ($4O$, 0.014 d) overtone radial modes. Only the $2O$ value has a
possible
counterpart in the frequency spectrum. i.e. the $f_5$ term. 
Figure~\ref{resi} shows the residual
signal after having considered the 12 frequencies as k.c.'s. We immediately note that no signal is detectable where
we
expect the $F$ mode: the noise in the 5--10 \cds region is
0.06 mmag and no peak is visible. The peak after 10 \cds is at 10.45 \cd,
which is not very close to the $1O$ value and thus its identification is
doubtful.  
The $3O$ value is very close to the orbital frequency of the
satellite and we cannot say anything about its possible presence.
Finally, the $4O$ mode is in a region where no signal is detectable.
It does not therefore seem that radial modes are excited to a level
comparable to 
that of the nonradial modes; in particular we note the absence
of the $F$ mode in a frequency region whose very low noise level would lead
us to
expect to detect it were it present.

Kennelly \& Walker (1996) reported spectroscopic observations of \td;
in addition to $f_1$, they also detected  
a high--degree mode at 16.0 \cd. Around that value, the noise level in our
residual
power spectrum is 0.12 mmag and no significant peak stands up.  
If the term reported by Kennelly \& Walker is real, 
then the lack of detectable amplitude variation
in the WIRE photometric series implies that 
cancellation effects are very effective on the integrated flux, thus
confirming the high $\ell$ degree  of this mode.
\subsection{The amplitude variability of modes}
The stability and the lifetime of the modes is an open point in
asteroseismology.
As \td was observed in the past, we can compare the previous results with
the new ones. Breger et al. (1987) identified the $f_3$, $f_6$, $f_1$ and
$f_2$ 
terms. Breger et al. (1989) added a fifth term, i.e. $f_9$. Li et al. (1997)
confirmed these five terms, but claimed evidence for amplitude variability
not
reported by Breger et al. (1989). 

It is immediately obvious that the relative strengths of the modes have
changed.
In the WIRE dataset
$f_1$ is by far the term with the largest amplitude, while it is only the
third-largest in Breger et al. (1989) and the fifth-largest in Li et al.
(1997). The largest amplitude term 
is $f_3$ in Breger et al. (1989)  and $f_6$ in Li et al. (1997). Note that
the
1~\cd alias interaction is  possible only between  $f_1$ and $f_{10}$ and, 
marginally, between $f_5$ and $f_7$. Taking also into account that the main
results from Breger et al. were obtained from 
a multisite campaign, the observed changes strengthen the hypothesis
of an amplitude variability rather than an interaction between aliases.

We also performed some simulations by introducing
an artificial drift of the $f_1$ amplitude to verify what threshold 
can originate a discernible  effect.  
We found that a spurious peak near $f_1$
appears for a linear drift as large as 0.08 mmag~d$^{-1}$, i.e for an 
amplitude variability attaining 12\% of the full amplitude of $f_1$.
Looking at Figure~\ref{resi} we can see that only a minor
peak is visible at 13.23~\cd and no peak is close to the highest amplitude
term $f_1$=13.697~\cd. The only pair suggestive of the presence of
amplitude variability is that composed of the $f_{12}$ and
$f_4$ terms (see Table~\ref{erro}). However, the relative amplitude
variability of the small amplitude $f_4$ term would have to be very large 
to produce such an effect, and that seems unlikely.
Therefore, we cannot infer any significant amplitude variability 
of the detected terms
over the 18.5--d baseline covered by the WIRE observations.
It should be noted that some $\delta$ Scuti stars do
display amplitude variability on this timescale (XX Pyx, Handler et al.
1998). 

\subsection {The frequency distribution}

\begin{figure}
\resizebox{\hsize}{!}{\includegraphics{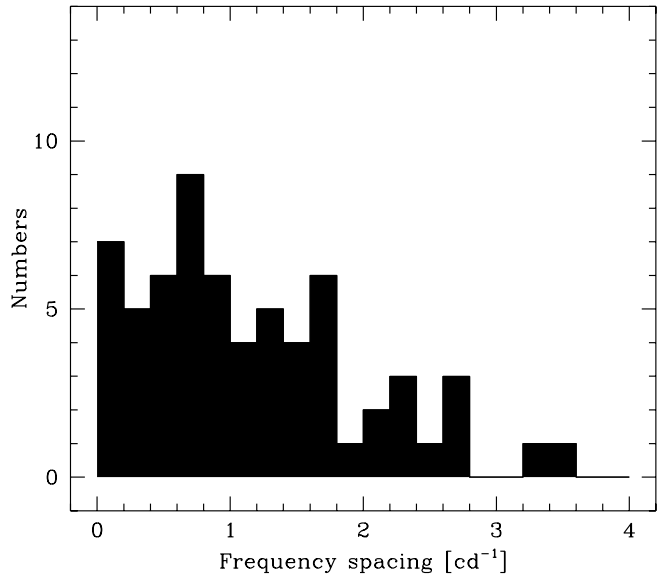}}
\caption[]{ Histograms of the frequency spacings between all the
frequency pairs (12 independent modes).  }
\label{isto}
\end{figure}

The frequency distribution of the modes
can be very different from one $\delta$ Scuti star to the next (see Figure~4
in Poretti 2000). 
\td displays a single bunch of frequencies, whose
average value makes \td more similar to 
to 4 CVn rather than XX Pyx; in any case, there is
no hint of two bunches of frequencies as in FG Vir. The investigation of
regularities in the frequency spacing distribution can supply details about
the stellar structure. Figure~\ref{isto} shows the histogram of the
differences
between all the frequency pairs; there is no particular peak, and
81\% of spacings are concentrated below 1.8~d. Below this limit, the
distribution is smooth; the more recurrent spacing is about 0.70~d.

Breger et al. (1989) concluded that the rotational splitting alone was not
able to explain the frequency spectrum they observed in the second multisite
campaign. They predicted that adjacent $m$ values would be separated by
\begin{equation}
\sigma_m - \sigma_{m+1} \geq 0.41~~ {\rm cd ^{-1}}
\end{equation}
This values is fixed at 0.50 \cds when considering i=45$^o$, i.e. the
rotational axis perpendicular to the orbital plane (Torres et al. 1997).
Although this separation is too large for the sample considered by Breger et
al. (1989)
it can now be detected in the more numerous frequency set detected by WIRE. 
A pulsational model of \td should therefore include a careful evaluation of
rotation; we will present this in a future paper.   
\section{Conclusions}
The results obtained on \td demonstrated the powerful capability
of a small instrument measuring stars from space, especially considering
that this particular use was unplanned.
In fact, the WIRE monitoring reported here puts \td among
the best studied $\delta$ Sct stars, i.e. among stars intensively observed
from ground by a large use of telescopes and manpower. 
The detection of the 12  terms having full--amplitude at the mmag level 
lowered the rms residual down to 1.5 mmag, i.e. about three times  the observational
error of the time series constituted by the 8958 normal points. Even admitting
other possible instrumental sources of errors,
that means that very probably undetected terms are again hidden in the light
curve; since they should have very small amplitude, they may be
very numerous. Therefore, $\delta$ Sct stars are confirmed as particularly
interesting targets for asteroseismology.
The interaction between spurious terms and signal is a
further complication: the spurious terms 
can be removed only on the basis of a step--by--step analysis and
a careful evaluation of their effect on the physical ones. 
In the specific case of \td, we had no problems with the $f$=26.19 \cd and $f<$1 \cd terms,
since they are far away from the frequencies where the signal is observed.
However,
$f$=15.05 \cd, i.e. the term introduced by the 
orbital period, masks from us a region where signal could be observed.

From a physical point of view, WIRE monitoring demonstrates us that \td is an 
interesting $\delta$ Sct star, showing numerous excited modes 
(and likely many more have not been detected yet) and amplitude 
variability. As we detected excited
terms only in a narrow interval, it is very probable that 
they originate from the primary component only.

The only drawback of the WIRE dataset is its relatively poor frequency
resolution
compared to ground-based multi-site efforts,
though this does not constitute a serious problem for the generally well separated
frequencies of \td.  However, close pairs of frequencies are observed
when going down to smallest amplitude: the requirement to achieve good frequency
resolution is essential to the success of future asteroseismological space missions. 
\begin{acknowledgements} 

We gratefully acknowledge the support of Harley Thronson, Phillipe Crane,
Daniel Golombek, and Joe
Bredekamp at NASA Headquarters for making this use of WIRE possible. The
hard work of many people,
including the WIRE operations and spacecraft teams at GSFC and the
timeline
generation team at IPAC,
was essential to the success of this project. While it is impractical to
single out everyone who
contributed, we would particularly like to thank Carol Lonsdale at IPAC
and
David Everett and
Patrick Crouse at GSFC for their efforts above and beyond the call of
duty.

\end{acknowledgements}

\end{document}